# On the Existence and Structure of Mixed Nash Equilibria for In-Band Full-Duplex Wireless Networks

Andrea Munari, Vaggelis G. Douros, Petri Mähönen

*Abstract*— This article offers the first characterisation of mixed Nash equilibria (MNE) for a wireless system with full-duplex capable terminals that share a common channel via Aloha-based contention. Focussing on a simple grid topology, we prove that mixed MNE exist only if proportionality, driven by the accuracy of self-interference cancellation, is granted between the cost undergone for full- and half-duplex operations. The analysis shows how a proper selection of such costs allows MNE that are also optimal from a network viewpoint in terms of aggregate throughput. The sensitivity of system performance to costs is tackled and discussed considering the price of anarchy.

## I. INTRODUCTION

In-band full-duplex (FD) is gaining momentum as an alternative paradigm for wireless communications. Simultaneous transmission and reception over the same band has indeed been proven viable for practical devices [1], unleashing alluring capacity improvements. In parallel, recent research efforts have brought a good understanding of the key tradeoffs FD triggers in more complex setups, buttressing the potential of the technique in settings ranging from ad hoc to cellular topologies [2], [3]. On the other hand, little attention has been devoted to the utility a single user within a network perceives when sending and receiving concurrently. In fact, FD operations might entail a higher cost, e.g. in terms of energy, complexity, or simply due to specific network pricing policies, possibly becoming unattractive. Understanding whether and under which conditions a user willingly embraces the new technology is thus paramount for its success.

The present article tackles this crucial yet still elusive question following a game-theoretic approach. To extract clear and fundamental insights we focus on a four-node topology, where players capable of transmitting in both half-duplex (HD) and FD mode share a common channel via slotted Aloha. As a primary contribution, we derive conditions for the existence of mixed Nash equilibria (MNE) without strictly dominated actions, highlighting the key role played by self-interference cancellation when employing FD. Under such constraints, we show that infinitely many MNE are admissible, characterise them analytically, and provide mechanism design hints on how to set the price of HD and FD transmissions to make them appealing from the single user's perspective. Moreover, the optimality of MNE from a network-wide perspective is discussed, proving that the maximum aggregate throughput can indeed be reached at an equilibrium point. The sensitivity of system performance to costs is also explored through the notion of price of anarchy.

Though medium access control (MAC) games have been studied in HD networks [4], our contribution is the first to apply them to Aloha-based FD systems, which are experiencing a revived interest for machine-type applications. In [5], FD communications are modelled focusing on coalitional games, whereas we tackle the non-cooperative case. Authors in [6], instead, consider a two-player game with FD and HD operation modes. Motivated by the potential inefficiency of a NE where both use FD, a Bayesian game is analysed, providing conditions for the existence and uniqueness of a Bayesian NE and discussing extensions to the multiple-player case. In contrast to this approach, this letter also considers the possibility for a node to wait instead of always transmitting using either FD or HD. Moreover, we analyse the existence and uniqueness of mixed rather than pure or Bayesian NE. The proper identification of the nodes' access probabilities is a key feature for the successful design and operation of a MAC protocol, and the study of MNE can help towards this direction. Finally, the proposed approach lays a base for the study of broader random access networks.

## II. SYSTEM MODEL AND PRELIMINARIES

Throughout this article, we focus on the grid topology reported in Fig. 1, composed of two node pairs $\mathcal{P}_i = (A_i, B_i)$, $i = 1, 2$. Users within a pair are located $r$ meters apart, while a distance $d = \kappa\, r$, $\kappa > 0$, separates the two couples. Time is divided in subsequent slots, whose duration equals the transmission time of a packet.[1] At every slot, each pair independently chooses how to access the shared wireless channel, following one strategy in the set $\mathcal{S} = \{w, t_A, t_B, t_{fd}\}$. When $w$ is selected, no transmission is performed by the pair over the slot. Conversely, $t_A$ indicates a HD data exchange from $A_i$ to $B_i$. In this case, the first node sends a packet to its mate, which acts as a pure receiver during the slot. Similarly, $t_B$ describes a link from $B_i$ to $A_i$. Finally, the choice of $t_{fd}$ implies the joint transmission of $A_i$ and $B_i$, concurrently sending data to each other resorting to FD techniques. From a different angle, the system instantiates a slotted Aloha access, where each node acts as a saturated source of packets for its pair-mate.[2]

All transmissions are performed with power $P$, and the wireless channel is affected by path-loss with exponent $\alpha$ and Rayleigh block fading. Accordingly, a sent packet reaches a node $x$ meters away from the source with power $P_r(x) = P\, x^{-\alpha} \zeta$, where $\zeta$ is an exponential r.v. of unit mean,

Authors are with the Institute for Networked Systems, RWTH Aachen University (Germany), e-mail: {firstname.lastname@inets.rwth-aachen.de}

---

[1]The terms *packet* and *data unit* will be used interchangeably.
[2]In this perspective, no feedback nor retransmission policies are considered.

independently drawn for each link and across slots. At the receiver side, decoding is hindered by additive noise of power $N$ and possibly by interference. The latter, in turn, is given by two components. Firstly, an undesired power $\mathcal{I}$ may reach the user from concurrent transmissions performed within the other pair. Secondly, if the node operates in FD mode, part of its emitted power will leak into the receiver chain as self-interference (SI), which we model following a well-established approach as $\eta P$, $\eta \in [0, 1)$ [2]. The overall signal to noise and interference ratio (SNIR) can thus be expressed as $P_r(r)/(N + \mathcal{I} + \eta P)$, where the last addend in the denominator is elided if a receiver is not concurrently transmitting. A threshold model for decoding is assumed, so that a data unit is retrieved at the intended addressee if and only if the experienced SNIR is larger than a parameter $\theta$, which embeds coding and modulation aspects.

Within this framework, the success probability of a transmission can be derived once the strategies $s_1, s_2 \in \mathcal{S}$ of the two pairs are given. Focussing on the reception of a packet at node $\mathsf{B}_i$, four cases have to be considered:

- $\mathcal{P}_j$ remains silent: no external interference affects reception at $\mathsf{B}_i$, and the success probability evaluates to

$$p_s(\mathsf{B}_i \,|\, s_j = w) = \mathbb{P}\left[\zeta > \frac{\theta N}{Pr^{-\alpha} + P\eta}\right]$$
$$= e^{-\theta(N+\eta P)r^\alpha/P}$$

leaning on the exponential distribution of the fading coefficient. The result can be conveniently written as $p_s(\mathsf{B}_i \,|\, s_j = w) = \beta\varphi$, where $\varphi := \exp(-\theta N r^\alpha/P)$ represents the success probability at a receiver in the sole presence of noise and $\beta := \exp(-\theta\eta r^\alpha)$ accounts for residual SI when the receiver operates in FD mode.[3]

- $\mathcal{P}_j$ chooses strategy $t_\mathsf{A}$: reception at $\mathsf{B}_i$ is affected by external interference coming from $\mathsf{A}_j$. Thus, $\mathcal{I} = P\zeta'(d^2 + r^2)^{-\alpha/2}$, where $\zeta'$ is the fading coefficient for the $\mathsf{A}_j$–$\mathsf{B}_i$ link. Conditioning on $\zeta'$ and recalling that $d = \kappa r$, the success probability evaluates to $\beta\varphi\exp(-\theta(\kappa + 1)^2\zeta')$. Averaging over the exponential distribution eventually leads to $p_s(\mathsf{B}_i \,|\, s_j = t_\mathsf{A}) = \beta\varphi\iota_f$, where the ancillary term

$$\iota_f := \frac{1}{1 + \theta(1 + \kappa^2)^{-\alpha/2}}$$

captures the effect at a receiver of the interference generated by the *farthest* node of the competing pair.

- $\mathcal{P}_j$ chooses strategy $t_\mathsf{B}$: reception at $\mathsf{B}_i$ is affected by external interference coming from $\mathsf{B}_j$. Proceeding along the steps discussed above, we readily get $p_s(\mathsf{B}_i \,|\, s_j = t_\mathsf{B}) = \beta\varphi\iota_c$, introducing

$$\iota_c := \frac{1}{1 + \theta\kappa^{-\alpha}}$$

to describe the impact at $\mathsf{B}_i$ of interference generated by the *closest* node of the other couple.

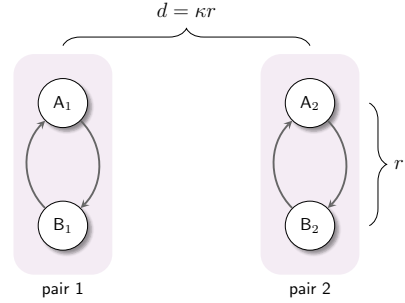

Fig. 1. Reference topology for the system under study.

- $\mathcal{P}_j$ chooses strategy $t_\mathsf{fd}$: in this case $\mathsf{B}_i$ is affected by interference coming from both $\mathsf{A}_j$ and $\mathsf{B}_j$. Recalling the mutual independence of all the fading coefficients, we obtain the sought success probability as $p_s(\mathsf{B}_i \,|\, s_j = t_\mathsf{fd}) = \beta\varphi\,\iota_c\iota_f$.

Due to symmetry, the derived formulations hold for a reception at node $\mathsf{A}_i$ as well.[4] Moreover, while the reported calculations have implicitly assumed the presence of self-interference ($\eta P$ term), the results also capture performance when nodes are operating in HD-mode, setting $\beta = 1$. For the remainder of our discussion, it is also useful to note that $0 < \iota_c < \iota_f < 1$ for any set of system parameters.

For the topology under study, we consider a normal form, simultaneous move, non-cooperative game with complete information, modelling each pair as a selfish rational player and assuming the set of players, the strategy sets and the utility functions to be common knowledge. More specifically, our focus is on mixed strategies, where at every slot a couple draws an action $s \in \mathcal{S}$ following the probability mass function (p.m.f.) $\boldsymbol{\pi} = (\pi_w, \pi_{t_\mathsf{A}}, \pi_{t_\mathsf{B}}, \pi_{t_\mathsf{fd}})$, independently of the behaviour of its competitor. To characterise the game, we introduce a utility function $U : \mathcal{S} \to \mathbb{R}^+$, which quantifies the average payoff earned by a player. We shall concentrate on utilities in the form $U(s) = \tau(s) - c(s)$, where $\tau(s)$ indicates the average throughput, defined as the expected number of packets successfully exchanged over the slot within the pair,[5] and $c(s)$ is the cost undergone to implement action $s$. As to the latter, we generalise a cost function extensively used in slotted Aloha with HD [4]: two distinct prices are foreseen for initiating a HD transmission ($c(t_\mathsf{A}) = c(t_\mathsf{B}) := c_\mathsf{hd}$) or a FD link ($c(t_\mathsf{fd}) := c_\mathsf{fd}$), whereas we set $c(w) = 0$. The definition is broadly applicable and captures, e.g. the cost to implement an action or the energy spent to transmit the packet. We remark how the choice of a throughput-based payoff aims to highlight a key tradeoff that emerges in FD networks. In fact, while the ability for both nodes in a pair to transmit simultaneously can potentially double the number of exchanged data units, it also leads to a higher aggregate level of interference, detrimental for the decoding probability. The blend of the two effects is captured by the very definition of $\tau(s)$.

---

[3]Throughout our analysis, we focus on $\beta > 1/2$. Such an assumption does not restrict the practical relevance of the derived results, as values of $\beta$ close to 1 are indeed required to leverage the potential of FD [2].

[4]Clearly, the success probability conditioned on $s_j = t_\mathsf{A}$ and $s_j = t_\mathsf{B}$ are swapped with respect to the ones experienced at $\mathsf{B}_i$.

[5]Note that the expectation that defines $\tau(s)$ for a player is with respect to both the fading coefficients for the involved links *and* the strategy followed by the competitor over the current slot, which drives the interference level $\mathcal{I}$.

To calculate the average throughput, let us focus without loss of generality on player $\mathcal{P}_i$. Clearly, if the node pair remains silent, $\tau(w) = 0$. Conversely, if a HD link is instantiated from $A_i$ to $B_i$, at most one data unit can be exchanged, and delivery is successful with probability $p_s(B_i|s_j)$. Averaging over the strategy of the competing player, we readily have

$$\tau(t_A) = \mathbb{E}_{s_j}[p_s(B_i\,|\,s_j)] = \varphi(\pi_w + \iota_f \pi_{t_A} + \iota_c \pi_{t_B} + \iota_c \iota_f \pi_{t_{fd}}).$$

Similarly, when strategy $t_B$ is picked

$$\tau(t_B) = \mathbb{E}_{s_j}[p_s(A_i\,|\,s_j)] = \varphi(\pi_w + \iota_c \pi_{t_A} + \iota_f \pi_{t_B} + \iota_c \iota_f \pi_{t_{fd}}).$$

Finally, if a FD connection is established, no packet is delivered (probability $[1 - p_s(A_i|s_j)][1 - p_s(B_i|s_j)]$); only one data unit gets through (probability $p_s(A_i|s_j)[1 - p_s(B_i|s_j)] + p_s(B_i|s_j)[1 - p_s(A_i|s_j)]$); or both packets contribute to the throughput (probability $p_s(A_i|s_j)p_s(B_i|s_j)$). Taking the average over $s_j$ leads after some manipulations to

$$\tau(t_{fd}) = \beta\varphi\big[2\pi_w + (\iota_c + \iota_f)\pi_{t_A} + (\iota_c + \iota_f)\pi_{t_B} + 2\iota_c \iota_f \pi_{t_{fd}}\big].$$

From these results, the average utility function $U(s)$ follows directly, and is reported for convenience in (1) at the bottom of the page.

## III. Mixed Nash Equilibria

For the system described in Sec. II, we aim at investigating the existence and structure of MNE without strictly dominated actions (i.e. so that for a player no action always yields a higher utility regardless of the one of the other player). This line of study allows to understand which medium sharing strategies in terms of access probabilities are appealing from a single user's perspective, leaning on game theory to derive MAC design insights. By definition, and considering the symmetry of the setting, we are thus interested in identifying a p.m.f. $\pi$ such that the average utility $U(s)$ of a player is the same regardless of the selected strategy $s \in \mathcal{S}$. A complete characterisation is offered by the following result:

*Theorem 1:* The considered topology can admit a MNE with no strictly dominated actions only if $\varphi \iota_c \iota_f \leq c_{hd} \leq \varphi$ and $c_{fd} = 2\beta c_{hd}$. Under these constraints, infinitely many MNE exist, obtained by picking $\pi_{t_{fd}}$ in the interval

$$\frac{\varphi(\iota_c + \iota_f) - 2c_{hd}}{\varphi(\iota_c + \iota_f - 2\iota_c\iota_f)} \leq \pi_{t_{fd}} \leq \frac{\varphi - c_{hd}}{\varphi(1 - \iota_c\iota_f)} \quad (2)$$

and setting

$$\begin{cases} \pi_w = \pi_{t_{fd}} \dfrac{\iota_c + \iota_f - 2\iota_c\iota_f}{2 - \iota_c - \iota_f} + \dfrac{2c_{hd} - \varphi(\iota_c + \iota_f)}{\varphi(2 - \iota_c - \iota_f)} \\ \pi_{t_A} = \pi_{t_B} = -\pi_{t_{fd}} \dfrac{1 - \iota_c\iota_f}{2 - \iota_c - \iota_f} + \dfrac{\varphi - c_{hd}}{\varphi(2 - \iota_c - \iota_f)} \end{cases} \quad (3)$$

*Proof:* To prove the necessary conditions, let us focus without loss of generality on action $t_A$, whose utility is reported in (1). Recalling that $\iota_c, \iota_f < 1$ and $\sum_{s \in \mathcal{S}} \pi_s = 1$, $U(t_A) \leq \varphi - c_{hd}$. Thus, for any $c_{hd} > \varphi$, action $t_A$ would be strictly dominated by action $w$ (i.e. $U(t_A) < U(w) = 0$, $\forall \pi$), in contrast with the sought MNE. Similarly, since $\iota_c \iota_f < \iota_c < \iota_f < 1$, the average throughput seen by a player selecting action $t_A$ is maximised for $\pi_{t_{fd}} = 1$. Then, $U(t_A) \geq \varphi \iota_c \iota_f - c_{hd}$ and $w$ would be strictly dominated by $t_A$ for any $c_{hd} < \varphi \iota_c \iota_f$. As to the second part of the theorem, we observe that, since $U(w) = 0$, a mixed strategy $\pi$ is a MNE iff $U(\pi_{t_A}) = U(\pi_{t_B}) = U(\pi_{t_{fd}}) = 0$ under the constraint $\sum_{s \in \mathcal{S}} \pi_s = 1$. From (1), this translates into the system of linear equations $\boldsymbol{A}\boldsymbol{\pi}^T = \boldsymbol{b}^T$, where

$$\boldsymbol{A} = \begin{pmatrix} \varphi & \varphi\iota_f & \varphi\iota_c & \varphi\iota_c\iota_f \\ \varphi & \varphi\iota_c & \varphi\iota_f & \varphi\iota_c\iota_f \\ 2\beta\varphi & \beta\varphi(\iota_c + \iota_f) & \beta\varphi(\iota_c + \iota_f) & 2\beta\varphi\iota_c\iota_f \\ 1 & 1 & 1 & 1 \end{pmatrix}$$

$\boldsymbol{b}$ is the row vector $(c_{hd}, c_{hd}, c_{fd}, 1)$, and $(\cdot)^T$ indicates the transpose operator. It is easy to verify that the third row of the coefficient matrix equals the sum of the first two multiplied by $\beta$. Thus, $\operatorname{rank}(\boldsymbol{A}) = 3$ and, by the Rouché-Capelli theorem, the system admits infinite solutions over a space of dimension 1 iff $c_{fd} = 2\beta c_{hd}$. Eqs. (2) and (3) follow after arithmetic manipulations by solving with respect to $\pi_w$, $\pi_{t_A}$ and $\pi_{t_B}$ under the constraints $\pi_w, \pi_{t_A}, \pi_{t_B} \in [0, 1]$. ∎

Theorem 1 offers relevant insights on the problem. In the first place, it highlights how a MNE can be reached only if the cost for establishing a FD link scales proportionally to that undergone for a HD one, i.e. $c_{fd} = 2\beta c_{hd}$. From this standpoint, the result pinpoints the key role of self-interference cancellation accuracy: the higher the residual SI level (i.e. the lower $\beta$), the lower the cost the system designer can charge for a FD link in order to reach an equilibrium and compensate for the worsened decoding performance. Secondly, the theorem reveals how the system can settle – for any set of network parameters – to one of infinitely many MNE involving FD transmissions (i.e. with $\pi_{t_{fd}} > 0$). Such an outcome is non-obvious and particularly appealing, as it confirms how one can indeed drive the players to an equilibrium that leverages the new technology by properly setting the costs. Further light on this aspect is shed by Fig. 2, which reports (shaded region) the values of $\pi_{t_{fd}}$ leading to a MNE when varying $c_{hd}$. As discussed, any cost outside the interval $[\varphi\iota_c\iota_f, \varphi]$ prevents the existence of equilibria without strictly dominated actions. More interestingly, the plot highlights how MNE with larger values of $\pi_{t_{fd}}$ are attainable only resorting to lower costs. Indeed, while for $\pi_{t_{fd}} = 0$ the cost can span between $\varphi(\iota_c + \iota_f)/2$ and $\varphi$,[6] medium access policies with $\pi_{t_{fd}}$ approaching 1 restrict $c_{hd}$ to its minimum (i.e. $\varphi\iota_c\iota_f$). This behaviour can be explained observing how players engaging more often in FD exchanges generate higher interference

---

[6]The former value follows by setting $\pi_{t_{fd}} = 0$ in (3) and by imposing the constraint $\pi_w \leq 1$.

$$\begin{aligned} U(w) &= 0 & U(t_A) &= \varphi\left(\pi_w + \iota_f\pi_{t_A} + \iota_c\pi_{t_B} + \iota_c\iota_f\pi_{t_{fd}}\right) - c_{hd} \\ U(t_B) &= \varphi\left(\pi_w + \iota_f\pi_{t_A} + \iota_c\pi_{t_B} + \iota_c\iota_f\pi_{t_{fd}}\right) - c_{hd} & U(t_{fd}) &= \beta\varphi\big[2\pi_w + (\iota_c + \iota_f)\pi_{t_A} + (\iota_c + \iota_f)\pi_{t_B} + 2\iota_c\iota_f\pi_{t_{fd}}\big] - c_{fd} \end{aligned} \quad (1)$$

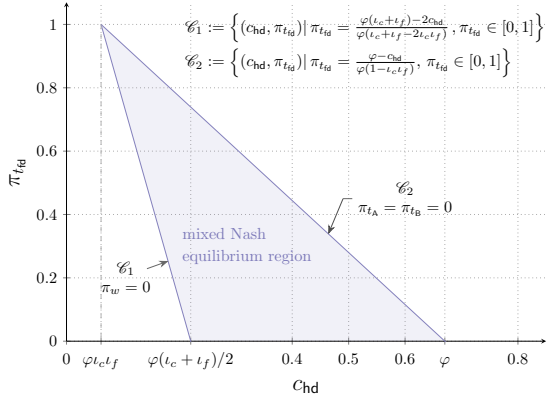 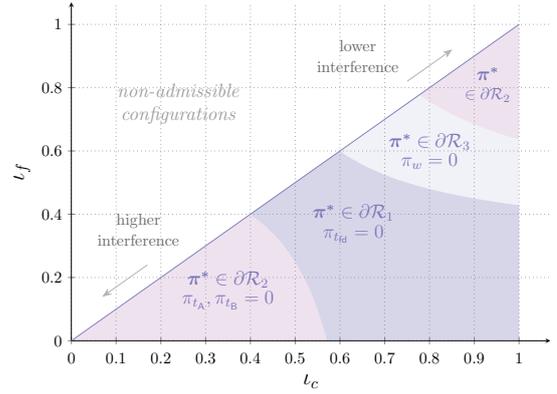

Fig. 2. The shaded region indicates values of $\pi_{t_{fd}}$ that lead to a MNE for a given $c_{hd}$. $\alpha = 3.5$, $\theta = 4$, $\kappa = 1$ and reference SNR $Pd^{-\alpha}/N = 10$.

Fig. 3. Maximum throughput configurations varying $\iota_c$ and $\iota_f$. $\beta = 0.7$.

which, in turn, lowers the success probability. A proportional reduction of the charging policies stands as the only way to enable MNE. From this viewpoint, Theorem 1 also offers a handy tool for system design, allowing to determine the costs of HD and FD links for the system to reach an equilibrium at a desired operating point in terms of medium access strategies. For the remainder of our discussion, it is useful to label the boundaries $\mathscr{C}_1$ and $\mathscr{C}_2$ of the equilibrium region, stated in (2) and reported for convenience in Fig. 2. Notably, inspection of (3) reveals how MNE over $\mathscr{C}_2$ involve no HD links (i.e. $\pi_{t_{hd}} = 0$), while the ones over $\mathscr{C}_1$ see players never refraining from accessing the channel ($\pi_w = 0$).

### A. On the Network Optimality of MNE

The possibility to drive the system to multiple equilibria involving any desired fraction of FD connections triggers the natural question of whether such operating conditions are also meaningful from a network perspective. Indeed, while the access strategy profile with $\pi_w = 1$ ($c_{hd} = \varphi$) is a MNE, it hardly represents a desirable working point. Similarly, an equilibrium with $\pi_{t_{fd}} = 1$ ($c_{hd} = \varphi\iota_c\iota_f$) might not be effective when the two players are close to each other, as the high level of mutual interference might lead to a very low packet delivery rate. We delve into this aspect by considering the aggregate throughput $\mathcal{T}_a$, defined as the mean number of packets per slot successfully exchanged within the four-node network. More specifically, the quantity depends on the actions undertaken by the players, and we are interested in the statistical average over a distribution $\boldsymbol{\pi}$, which we constrain for the sake of fairness to satisfy $\pi_{t_A} = \pi_{t_B} := \pi_{t_{hd}}$ (i.e. $\pi_w + 2\pi_{t_{hd}} + \pi_{t_{fd}} = 1$). The computation of $\mathcal{T}_a$ requires the inspection of all 16 players' actions combinations, which we report in Appendix. For each of them, in turn, the two-pair throughput follows from suitably applying the approach of Sec. II. Simple arithmetic eventually leads to a compact expression in $\pi_{t_{hd}}$ and $\pi_{t_{fd}}$:

$$\mathcal{T}_a = 4\varphi(\pi_{t_{hd}} + \beta\pi_{t_{fd}})\bigl[1 - \pi_{t_{hd}}(2 - \iota_c - \iota_f) - \pi_{t_{fd}}(1 - \iota_c\iota_f)\bigr]. \quad (4)$$

Despite the clean structure of (4), finding the maximum throughput $\mathcal{T}_a^*$ in closed-form is not trivial. We observe instead how the problem is to be solved within the closed region $\mathcal{R} \subset \mathbb{R}^2$, with boundary $\partial\mathcal{R} = \partial\mathcal{R}_1 \cup \partial\mathcal{R}_2 \cup \partial\mathcal{R}_3$ defined by the segments $\partial\mathcal{R}_1 = \{(\pi_{t_{hd}}, \pi_{t_{fd}})|\pi_{t_{hd}} \in [0, 1/2], \pi_{t_{fd}} = 0\}$, $\partial\mathcal{R}_2 = \{(\pi_{t_{hd}}, \pi_{t_{fd}})|\pi_{t_{hd}} = 0, \pi_{t_{fd}} \in [0, 1]\}$ and $\partial\mathcal{R}_3 = \{(\pi_{t_{hd}}, \pi_{t_{fd}})|\pi_{t_{hd}} \in [0, 1/2], \pi_{t_{fd}} = 1 - 2\pi_{t_{hd}}\}$. If we now take the partial derivatives of $\mathcal{T}_a$ over $\pi_{t_{hd}}$ and $\pi_{t_{fd}}$, it is possible to verify that the function has a unique critical point, which falls out of $\mathcal{R}$ for all parameters configurations. Thus, the maximum is found for a p.m.f. $\boldsymbol{\pi}^*$ that lies on $\partial\mathcal{R}$, where $\mathcal{T}_a$ becomes a tractable function of a single variable. This brings a first interesting remark, as we infer that the peak throughput is always achieved having at least one of the medium access probabilities $\pi_w$, $\pi_{t_{hd}}$, or $\pi_{t_{fd}}$, set to zero. Moreover, the original problem reduces to identifying the highest among the maxima exhibited by $\mathcal{T}_a$ over the three segments composing $\partial\mathcal{R}$, whose values are reported in (5). Which of the three actually coincides with $\mathcal{T}_a^*$ depends on the specific parameters of the network. From this standpoint, while conceptually simple, an exact characterisation of all possible configurations would be cumbersome. More interestingly, we introduce the following result:

*Theorem 2:* The topology under consideration always admits a MNE with no strictly dominated actions that reaches the maximum aggregate throughput.

*Proof:* It is sufficient to show that all the p.m.f. leading to a maximum of $\mathcal{T}_a$ (reported in (5)) indeed instantiate a MNE. Assume first that maximum throughput is achieved over $\partial\mathcal{R}_2$, so that $\pi_{t_{hd}} = 0$. In this case, the result is proven recalling the discussion of Fig. 2. In fact, for any $\pi_{t_{fd}}$, a MNE does exist over the boundary $\mathscr{C}_2$, where $\pi_{t_{hd}} = 0$. Similarly, any configuration obtained moving along $\partial\mathcal{R}_3$ (i.e. $\pi_w = 0$) corresponds to a MNE over the boundary $\mathscr{C}_1$ of the equilibria region. Finally, assume that $\mathcal{T}_a^*$ is achieved on $\partial\mathcal{R}_1$. From (5), when $\iota_c + \iota_f \leq 1$, throughput-maximisation requires $\pi_{t_{fd}} = 0$ and $\pi_{t_{hd}} = 1/[2(2 - \iota_c - \iota_f)]$. Plugging these values into (3) shows how the p.m.f. indeed represents a MNE, achieved setting $c_{hd} = \varphi/2$. Likewise, for $\iota_c + \iota_f > 1$, $\mathcal{T}_a^*$ is reached for $\pi_{t_{fd}} = 0$ and $\pi_{t_{hd}} = 1/2$, which is also, from (3), a MNE when $c_{hd} = \varphi(\iota_c + \iota_f)/2$. ∎

We then conclude that the system always admits a MNE that is also optimal from a network perspective by a suitable choice of the policy costs. Additional insights are captured by Fig. 3, which reports by different colours the access conditions leading to the peak aggregate throughput. Results were obtained by means of numerical evaluation of (5) for any possible configurations of system parameters embedded by $\iota_c$ and $\iota_f$, and assuming $\beta = 0.7$. As expected, when the interference level between the pairs is low ($\iota_c, \iota_f \sim 1$), e.g. due to a large distance between them, an aggressive solution where nodes either wait or transmit in FD mode yields the largest throughput. Increasing the impact of the mutual disturbance, the optimal point starts embedding also HD links, up to the point where giving up FD completely becomes convenient ($\pi_{t_{\text{fd}}} = 0$). Finally, when the interference level is severe (low $\iota_c$ and $\iota_f$), the best solution is again to defer access – with high probability – or to be aggressive (FD link) in the hope that both packets will get through if the competitor waits.

While the derived result is remarkable, proper mechanism design shall recall that changes in the cost policies may lead to distinct MNE with different performance. A relevant question is thus how sensitive $\mathcal{T}_a$ is to variations in $c_{\text{hd}}$. The issue is tackled in Fig. 4, which reports for any value of $\pi_{t_{\text{fd}}}$ the corresponding price of anarchy (PoA), defined as the ratio of $\mathcal{T}_a^*$ to the minimum aggregate throughput over all possible MNE [7]. Two settings are depicted, instantiating high ($\iota_c = 0.1$, $\iota_f = 0.2$) and low ($\iota_c = 0.6$, $\iota_f = 0.7$) cross-pair interference. As expected, the PoA diverges for $\pi_{t_{\text{fd}}} \to 0$, since in these conditions the null-throughput MNE $\pi_w = 1$ ($c_{\text{hd}} = \varphi$) becomes admissible (see Fig. 2). More interestingly, the plot pinpoints how care is required in setting the costs under harsh interference conditions, e.g. due to pairs that are close to each other. Deviations from the optimal pricing policy lead in fact to severe losses in terms of throughput, all the more so when nodes attempt FD connections. The issue eases considerably under lighter interference, where a large fraction of MNE offer similar (and close to optimal) performance. In such conditions, this desirable feature is further enhanced by improving SI cancellation capabilities ($\beta = 0.9$, dashed line). The effect is reversed for severe cross-pair disturbance, when lower SI levels lead to an increase of the peak throughput but are not enough to boost the worst performance among all MNE.

## IV. Conclusions and Future Work

Focussing on a 4-node topology with FD-capable terminals and Aloha-based medium access, this letter characterised the existence and structure of MNE. It was shown that an equilibrium optimal also from a networking viewpoint always exists. Commented results offer first design insights, and are meant to further stimulate game-theoretic research for FD networks, considering broader ad hoc topologies. Future work includes a study of pure NE as well as of algorithms leading the system to a desired equilibrium.

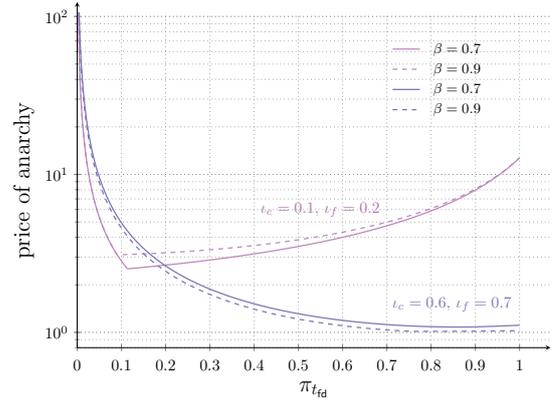

Fig. 4. Price of anarchy vs $\pi_{t_{\text{fd}}}$ for different values of $\beta$.

## Appendix I
### Derivation of the Aggregate Throughput

Following the definition of Sec. III, the average aggregate throughput can be expressed as

$$\tau_n = \sum_{\bm{s} \in \mathcal{S}^2} \mathbb{P}[\,(s_1, s_2) = \bm{s}\,] \, \tau(\bm{s}) \qquad (6)$$

where $\tau(\bm{s})$ is the number of packets delivered over a slot within the two pairs given profile $\bm{s}$, averaged over the distribution of the involved fading coefficients. The calculation of $\tau_n$ can then simply proceed by inspection of the 16 possible strategy combinations, leaning on the derivation of the success probability in Sec. II and recalling the independence of the fading coefficients. In details:

- the profile $(w, w)$, selected with probability $\pi_w^2$, clearly brings no contribution to the throughput;
- 4 profiles involve the transmission of a single node, while all others remain silent. Each such profile occurs with probability $\pi_w \pi_{t_{\text{hd}}}$, and leads to $\tau(\bm{s}) = \varphi$ (at most one packet is delivered, with probability $\varphi$);
- 2 profiles, each of probability $\pi_{t_{\text{hd}}}^2$, see the transmission of the two closest nodes (i.e. either $A_1$ and $A_2$, or $B_1$ and $B_2$). In this situation, one packet may go through while the other fails (overall probability $2\varphi\iota_c(1 - \varphi\iota_c)$) or both nodes may succeed and two data units be delivered (probability $(\varphi\iota_c)^2$). We thus get $\tau(\bm{s}) = 2\varphi\iota_c(1 - \varphi\iota_c) + 2(\varphi\iota_c)^2 = 2\varphi\iota_c$;
- 2 profiles see the transmission of the two farthest nodes (i.e. $A_1$ and $B_2$, or $B_1$ and $A_2$). Following the argument just described, each has probability $\pi_{t_{\text{hd}}}^2$ and $\tau(\bm{s}) = 2\varphi\iota_f$;
- 2 profiles involve the FD transmission of one pair, while the other remains silent (probability $\pi_w \pi_{t_{\text{fd}}}$). Both packets are successful with probability $(\beta\varphi)^2$, while a single success occurs with overall probability $2\beta\varphi(1-\beta\varphi)$. The average throughput follows as $\tau(\bm{s}) = 2\beta\varphi$;

$$\begin{aligned}
\partial\mathcal{R}_1 &: \mathcal{T}_a^* = \varphi/(2 - \iota_c - \iota_f) & \pi_{t_{\text{fd}}} &= 0 & \pi_{t_{\text{hd}}} &= \min\{1/2, 1/[2(2 - \iota_c - \iota_f)]\} \\
\partial\mathcal{R}_2 &: \mathcal{T}_a^* = \beta\varphi/(1 - \iota_c\iota_f) & \pi_{t_{\text{fd}}} &= \min\{1, 1/[2(1 - \iota_c\iota_f)]\} & \pi_{t_{\text{hd}}} &= 0 \\
\partial\mathcal{R}_3 &: \mathcal{T}_a^* = \frac{\varphi\big[\iota_c\iota_f\beta(\iota_c + \iota_f)\big]^2}{(2\beta - 1)(\iota_c + \iota_f - 2\iota_c\iota_f)} & \pi_{t_{\text{fd}}} &= \frac{(1 - \beta)(\iota_c + \iota_f) - \iota_c\iota_f}{(2\beta - 1)(2\iota_c\iota_f - \iota_c - \iota_f)} & \pi_{t_{\text{hd}}} &= (1 - \pi_{t_{\text{fd}}})/2
\end{aligned} \qquad (5)$$

- with probability $\pi_{t_{\text{fd}}}\pi_{t_{\text{hd}}}$, one player initiates a FD link while the other attempts a HD communication (4 possible profiles). All links (i.e. 3 packets) succeed with probability $(\beta\varphi)^2\iota_c\iota_f \cdot \varphi\iota_c\iota_f$. Two data units are exchanged if both nodes of the FD pair decode while the HD link fails (probability $(\beta\varphi)^2\iota_c\iota_f \cdot (1-\varphi\iota_c\iota_f)$) or if the HD link succeeds and one of the FD connection does not (probability $\beta\varphi\iota_c(1-\beta\varphi\iota_f)\cdot\varphi\iota_c\iota_f + \beta\varphi\iota_f(1-\beta\varphi\iota_c)\cdot\varphi\iota_c\iota_f$). Finally, a single packet is delivered if either the FD connection fails while the HD receiver decodes (probability $(1-\beta\varphi\iota_c)(1-\beta\varphi\iota_f)\cdot\varphi\iota_c\iota_f$), or if one of the FD-operated nodes succeeds and all others do not (probability $\beta\varphi\iota_c(1-\beta\varphi\iota_f)\cdot(1-\varphi\iota_c\iota_f)+\beta\varphi\iota_f(1-\beta\varphi\iota_c)\cdot(1-\varphi\iota_c\iota_f)$). Combining the enumerated possibilities, brief manipulations lead to $\tau(s)=\varphi\big[\beta(\iota_c+\iota_f)+\iota_c\iota_f\big]$;
- with probability $\pi_{t_{\text{fd}}}^2$ both players instantiate a FD connection. Simple arguments show that: 4 packets are decoded with probability $(\beta\varphi\iota_c\iota_f)^4$; 4 combinations, each of probability $(\beta\varphi\iota_c\iota_f)^3\cdot(1-\beta\varphi\iota_c\iota_f)$, bring delivery of 3 data units; 6 configurations allow retrieval of 2 packets (probability $(\beta\varphi\iota_c\iota_f)^2\cdot(1-\beta\varphi\iota_c\iota_f)^2$); and 4 combinations are to be considered for a single success, with probability $\beta\varphi\iota_c\iota_f\cdot(1-\beta\varphi\iota_c\iota_f)^3$. The throughput of the $(t_{\text{fd}},t_{\text{fd}})$ profile thus follows as $4\beta\varphi\iota_c\iota_f$.

Plugging the enumerated cases into (6) leads to (4) after basic arithmetic manipulations.